 \documentclass[twocolumn,prl,aps,psfig,showpacs,preprintnumbers,superscriptaddress]{revtex4}
\usepackage{graphicx}
\usepackage{epstopdf}
\usepackage{float}
\usepackage{color}
\usepackage{amsmath}

\newcommand{\feses}{FeSe$_{0.91}$S$_{0.09}$}

\newcommand{\fesex}{FeSe$_{1-x}$S$_{x}$}

\begin{document}

\title{Impact of Nematicity on the Relationship between Antiferromagnetic Fluctuations and Superconductivity in FeSe$_{0.91}$S$_{0.09}$ Under Pressure}

\author{K. Rana}
\affiliation{Ames Laboratory, U.S. DOE, and Department of Physics and Astronomy, Iowa State University, Ames, Iowa 50011, USA}
\author{Li. Xiang}
\affiliation{Ames Laboratory, U.S. DOE, and Department of Physics and Astronomy, Iowa State University, Ames, Iowa 50011, USA}
\author{P. Wiecki}
\affiliation{Ames Laboratory, U.S. DOE, and Department of Physics and Astronomy, Iowa State University, Ames, Iowa 50011, USA}
\affiliation{Karlsruhe Institute of Technology, Institut f\"ur Festk\"orperphysik, 76021 Karlsruhe, Germany}
\author{R. A. Ribeiro}
\affiliation{Ames Laboratory, U.S. DOE, and Department of Physics and Astronomy, Iowa State University, Ames, Iowa 50011, USA}
\author{G. G. Lesseux}
\affiliation{Ames Laboratory, U.S. DOE, and Department of Physics and Astronomy, Iowa State University, Ames, Iowa 50011, USA}
\author{A. E. B\"ohmer}
\affiliation{Ames Laboratory, U.S. DOE, and Department of Physics and Astronomy, Iowa State University, Ames, Iowa 50011, USA}
\affiliation{Karlsruhe Institute of Technology, Institut f\"ur Festk\"orperphysik, 76021 Karlsruhe, Germany}
\author{S. L. Bud'ko}
\affiliation{Ames Laboratory, U.S. DOE, and Department of Physics and Astronomy, Iowa State University, Ames, Iowa 50011, USA}
\author{P. C. Canfield}
\affiliation{Ames Laboratory, U.S. DOE, and Department of Physics and Astronomy, Iowa State University, Ames, Iowa 50011, USA}
\author{Y. Furukawa}
\affiliation{Ames Laboratory, U.S. DOE, and Department of Physics and Astronomy, Iowa State University, Ames, Iowa 50011, USA}
\date{\today}

\begin{abstract}

   The sulfur substituted FeSe system, FeSe$_{1-x}$S$_{x}$, provides a versatile platform  for studying the relationship between nematicity, antiferromagnetism, and superconductivity.  Here, by nuclear magnetic resonance (NMR) and resistivity measurements up to 4.73 GPa on FeSe$_{0.91}$S$_{0.09}$, we established the pressure($p$)-temperature($T$) phase diagram in which  the nematic state is suppressed with pressure showing a nematic quantum phase transition (QPT) around $p$ = 0.5 GPa,  two  SC regions, separated by the QPT, appear and antiferromagnetic (AFM) phase emerges above $\sim$3.3 GPa.   From the NMR results up to 2.1 GPa, AFM fluctuations are revealed to be characterized by the stripe-type wavevector which remains the same for the two SC regions.  Furthermore, the electronic state is found to change in character from non-Fermi liquid  to Fermi liquid  around the nematic QPT and persists up to $\sim$ 2.1 GPa.    In addition, although the AFM fluctuations correlate with $T_{\rm c}$ in both SC states, demonstrating the importance of the AFM fluctuations for the appearance of  SC in the system,  we found that, when nematic order is absent, $T_{\rm c}$ is strongly correlated with the AFM fluctuations,  whereas  $T_{\rm c}$ weakly depends on the  AFM fluctuations when nematic order is present.   Our findings on FeSe$_{0.91}$S$_{0.09}$  were shown to be applied to the whole FeSe$_{1-x}$S$_{x}$ system and also provide a new insight into the relationship between AFM fluctuations and SC in Fe-based superconductors.

\end{abstract}
\maketitle 

\begin{figure}[tb]
\includegraphics[width=\columnwidth]{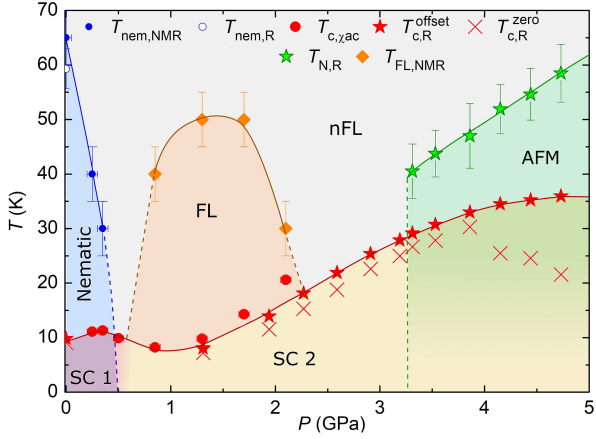} 
\caption{$p-T$ phase diagram of \feses.  
   The nematic transition temperatures $T_{\rm nem,NMR}$ and $T_{\rm nem,R}$ are determined by the splitting of the NMR spectrum under $H||ab$ and resistivity measurements at $H$ = 0, respectively. 
 $T_{\rm{c,\chi_{ac}}}$ (red solid circles) denotes $T_{\rm c}$ under zero magnetic field,  determined by  ${\it in~situ}$ ac susceptibility measurements using the NMR coil.
 $T_{\rm c, R}^{\rm zero}$ (red stars)  and $T_{\rm c, R}^{\rm offset}$ (crosses) denote $T_{\rm c}$ at $H$ = 0  determined  by zero resistivity and offset point, respectively,  in resistivity measurements \cite{supp}.  
 The AFM transition temperature ($T_{\rm N}$) was determined by resistivity measurements  \cite{supp}. 
  $T_{\rm FL, NMR}$ represents a crossover temperature between non-Fermi liquid (nFL) and Fermi liquid (FL) states determined by 1/$T_1$ measurements: Curie-Weiss like behavior of 1/$T_1T$ for nFL  and 1/$T_1T$ = constant Korringa behavior for FL. 
The solid and dotted lines are guides for the eyes.
}
\label{fig:Fig1}
\end{figure}  

    The interplay between magnetic fluctuations, electronic nematicity and the unconventional nature of superconductivity (SC) has received wide interest after the discovery of high $T_{\rm c}$ SC in iron pnictides \cite{Kamihara2008}.
    In most of the iron pnictide superconductors, by lowering temperature, the crystal structure changes from high-temperature tetragonal (HTT, $C_4$ symmetry)  to low-temperature orthorhombic (LTO, $C_2$ symmetry) at, or just above, a system-dependent N\'eel temperature $T_{\rm N}$, below which long-range stripe-type antiferromagnetic (AFM) order emerges \cite{Canfield2010,Johnston2010,Stewart2011,Scalipino2013}.
   SC in these compounds emerges upon suppression of both the structural (or nematic) and magnetic transitions by carrier doping and/or the application of pressure ($p$).
    While this clearly suggests a close relationship between AFM and nematic phases, the individual contribution to SC from these two phases becomes difficult to separate.

    In this context, the sulfur substituted FeSe system, \fesex, provides a favorable platform for the study of the impact of nematicity or antiferromagnetism on SC independently \cite{Bohmer2018}. 
    The superconductor FeSe ($x$ = 0) with a critical temperature of $T_{\rm c}$ = 8.5 K exhibits only a HTT-LTO structural phase transition, corresponding to a nematic phase transition, at $T_{\rm nem}$ = 90 K without AFM ordering  under ambient pressure \cite{Hsu2008,McQueen2009,Bohmer2018}.  
    With increasing $x$, the nematic phase is suppressed  and a nematic quantum phase transition (QPT) was reported to be around $x$ = 0.17 \cite{Hosoi2016}.  
    In contrast, $T_{\rm c}$ first increases from $T_{\rm c}$ = 8.5 K up to 10 K around $x$ = 0.09 \cite{Abdel2015,Watson2015,Reiss2017}, 
then is suppressed at higher $x$, whereas the fully replaced FeS is still a superconductor with $T_{\rm c}$ = 5 K \cite{Lai2015}. 
     As in the case of FeSe, no AFM state has been observed in \fesex~at ambient pressure, making this a suitable system to study the effects of nematicity on SC \cite{Abdel2015,Watson2015,Reiss2017}.
      Spectroscopic-imaging scanning tunneling microscopy \cite{Hanaguri2018}, thermal conductivity and specific heat \cite{Sato2018} showed that the gap anisotropy and its size change drastically at the nematic QPT. 
     Shubnikov-de Haas oscillation measurements indicate a change in both the topology of the Fermi surface and the degree of electronic correlations across the nematic QPT \cite{Coldea2019}. 
     These results suggest that the presence or absence of nematicity result in two distinct superconducting states. 
      Although no AFM state is observed in \fesex~under ambient pressure, the correlations between $T_{\rm c}$ and AFM fluctuations have been pointed out from nuclear magnetic resonance (NMR) measurements \cite{Wiecki2018, Beak2020}. 
   
    With the application of pressure on \fesex, the nematic state can also be suppressed and an AFM state is induced \cite{Xiang2017,Matsuura2017}. 
    The  three dimensional $T-p-x$ phase diagram up to $p$ = 8 GPa has been reported by Matsuura {\it et al.} \cite{Matsuura2017} in which the AFM ordered phase  shifts to higher $p$ with increasing $x$, although a different phase diagram of  FeSe$_{0.89}$S$_{0.11}$ having a wide AFM region was recently reported  \cite{Holenstein2019}.
     Recent resistivity measurements  under high magnetic fields on FeSe$_{0.89}$S$_{0.11}$ under pressure reported a lack of nematic quantum criticality and the presence of Fermi-liquid behavior \cite{Reiss2019}.  
     In addition,  two SC domes separated by the nematic QPT under magnetic field have been reported in \fesex~with $x$ = 0.12 \cite{Kuwayama2019} and 0.11 \cite{Reiss2019} under pressure, which was not reported in the first phase diagram \cite{Matsuura2017}.    
     To clarify this, it is crucial to establish the $p-T$ phase diagram and also to investigate the change in the character of AFM fluctuations and its relationship with SC across a nematic QPT in \fesex~under pressure.

     In this paper, we have carried out NMR and resistivity measurements on \feses~under pressure to investigate its physical properties from a microscopic point of view, especially focusing on  the differences in the AFM fluctuations between the two different SC domes and their  relationship with $T_{\rm c}$.    
   Based on the present NMR and resistivity data \cite{supp}, we established the phase diagram as a function of $p$ shown in Fig.~\ref{fig:Fig1}. 
    Similar to the case of $x$ = 0.11 and 0.12, a double SC dome structure is observed.
    From the temperature dependence of nuclear spin-lattice relaxation rate (1/$T_1$), we found a crossover from non-Fermi-liquid (nFL) to Fermi-liquid (FL) states with pressure and a dome-shaped FL phase between nematic and AFM phases.   
    In addition, although we inferred  that the wavevector of AFM fluctuations is stripe type for both superconducting domes and does not change with pressure, the  symmetry ($C_4$ or $C_2$) of the AFM fluctuations has been revealed to play an important role for superconducting transition temperature.

      Single crystals  of \feses~were prepared using vapor transport method as outlined in Ref. \cite{Bohmer2016}. 
    The details of the single crystals used for NMR measurements were described in Ref. \cite{Wiecki2018}.
    NMR measurements of $^{77}$Se nuclei  ($I$ = 1/2, $\gamma_{\rm N}$/2$\pi$ = 8.1432 MHz) under  a fixed magnetic field $H$ = 7.4089 T  \cite {NMR}
have been carried out  by using a lab-built spin-echo spectrometer up to a pressure of 2.10 GPa with a NiCrAl/CuBe piston-cylinder cell using  Daphne 7373 as the pressure transmitting medium. 
    Pressure calibration was accomplished by $^{63}$Cu nuclear quadruple resonance in Cu$_2$O \cite{Fukazawa2007,Reyes1992} at 77 K.
    Resistivity measurements under higher pressures up to 4.73 GPa were carried out in a modified Bridgeman anvil type cell \cite{Colombier2007} using  a 1:1 mixture of iso-pentane:n-pentane as the pressure medium.

\begin{figure}[tb]
\includegraphics[width=\columnwidth]{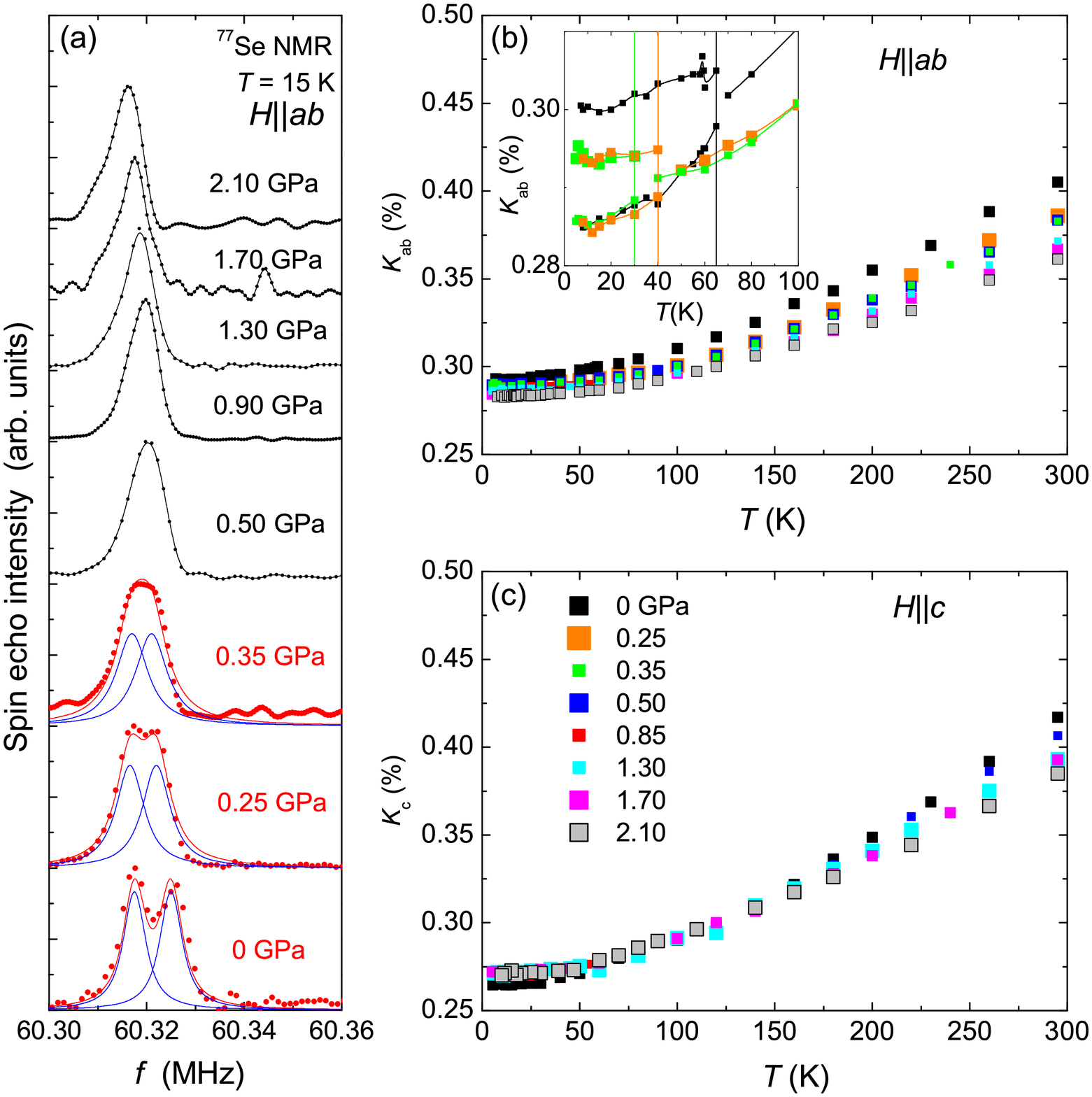} 
\caption{(a) Pressure dependence of $^{77}$Se NMR spectra of \feses~at 15 K for $H||ab$. 
        Below 0.50 GPa, the clear double peak structures (shown in red) are observed due to nematic phase transition, which can be well reproduced by the two Lorentzian curves shown in blue. 
     (b) Temperature dependence of $^{77}$Se NMR Knight Shift ($K$) at various pressures with $H||ab$. When splitting of line was present in the nematic state, the average values of $K$ were plotted. 
     The inset shows $K$ values for the split lines below the nematic temperatures. (c) $K$ for all measured pressures with $H||$c. For this $H$ direction, no splitting of spectra was observed.}
\label{fig:Fig2}
\end{figure} 

\begin{figure*}[tb]
\includegraphics[width=2\columnwidth]{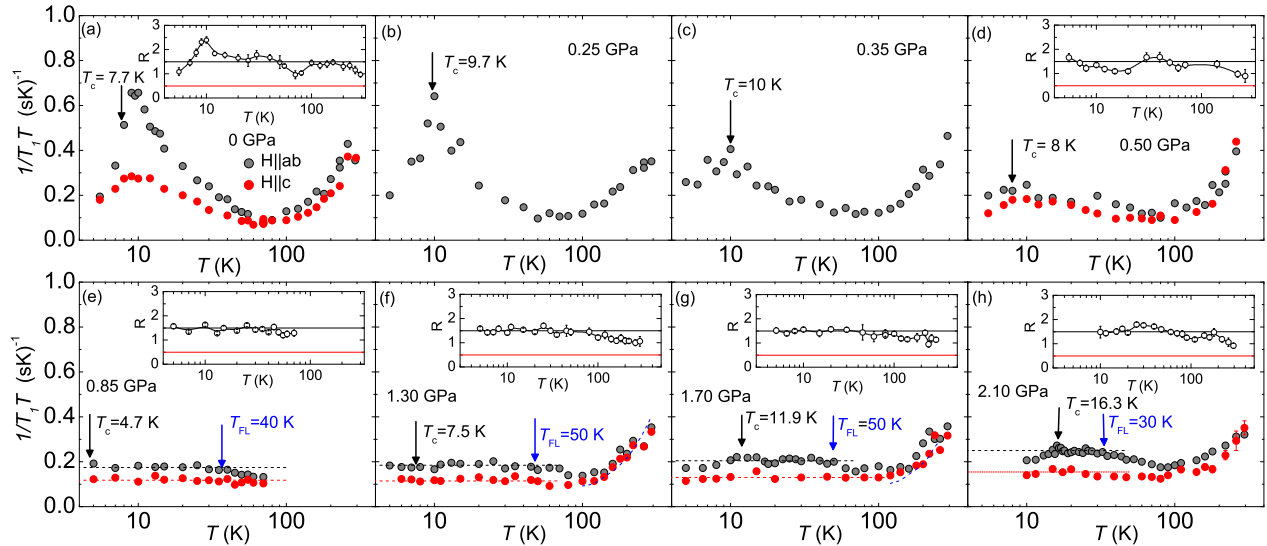} 
\caption{Temperature dependence of $^{77}$Se NMR 1/$T_1T$  at various pressures with $H||ab$ (gray circles) and $H||c$ (red circles). 
    Black arrows show $T_{\rm c}$ under $H||ab$ = 7.4089 T determined by the ${\it in~situ}$ ac susceptibility measurements. 
   Blue arrows show the temperature below which $1/T_1T$ = constant behavior is observed, defined as $T_{\rm FL}$. 
The inset of each panel shows the temperature dependence of  the ratio $R$ $\equiv$ $(1/T_1T)_{ab}$/$(1/T_1T)_{c}$. 
    The two horizontal lines represent the expected values for stripe-type ($R$ = 1.5) and  N\'eel-type ($R$ = 0.5) AFM fluctuations, respectively.  
}
\label{fig:Fig3}
\end{figure*}

  Figure \ref{fig:Fig2}(a) shows the $^{77}$Se NMR spectra of \feses~measured at 15 K under various pressures ($p$ = 0 - 2.10 GPa) with $H$ parallel to the $ab$ plane ($H||ab$).
   Here we applied magnetic field along [110] direction in the HTT phase. 
   As reported in Ref. [\onlinecite{Wiecki2018}], a clear splitting of the line due to  nematic order is observed at ambient pressure below $T_{\rm nem}$ $\sim$ 60 K \cite{supp}. 
   Although the splitting becomes small with increasing $p$,  the two-peak structure can be observed up to 0.35 GPa as shown in red in Fig.~\ref{fig:Fig2}(a) where the spectra are well reproduced by the sum of two peaks shown in blue, evidencing the nematic order up to 0.35 GPa. 
    On the other hand, no clear splitting of the line can be observed above 0.5 GPa. 
   Even at $T$ = 4 K, we do not observe the splitting, indicating no nematic order above 0.5 GPa. 
   From the smooth extrapolation of the $p$ dependence of $T_{\rm nem}$ described below [also, see Fig. 1], we found a nematic QPT around 0.5 GPa in \feses.

     Figures~\ref{fig:Fig2}(b) and \ref{fig:Fig2}(c) show the temperature dependence of the Knight shift ($K$) for $H||ab$  and  $H$ parallel to the $c$ axis ($H||c$), respectively. 
     The inset in Fig. ~\ref{fig:Fig2}(b) shows two values of $K$ for the two peaks observed  in the nematic state, from which $T_{\rm nem}$ is determined to be $\sim$65 K, $\sim$40 K, and $\sim$30 K for ambient, 0.25, and 0.35 GPa, respectively. 
   The estimated values of $T_{\rm nem}$ are consistent with the previous report \cite{Xiang2017}.
     In the main panel of Fig. \ref{fig:Fig2}(b), the average values of $K$ for the two peaks were plotted. 
    When $H||c$, no splitting of the line was observed.     
    Throughout all pressures and both $H$ directions, the values of $K$  are nearly independent of $p$, although $K$ seems to be suppressed very slightly with $p$ \cite{K_details}. 
    As shown, $K$ values are nearly constant below $\sim$50 K and then increase  with temperature above 100 K. 
    The nearly $p$ independent behavior of $K$ indicates that static uniform magnetic susceptibility is nearly independent of $p$, especially at low temperatures.  
     This also suggests that the application of pressure up to 2.10 GPa  does not produce significant change in the density of states at the Fermi energy $N(E_{\rm F})$ \cite{K data}, even though $T_{\rm c}$ varies significantly.
      This is in contrast to conventional BCS superconductors, in which $N(E_{\rm F})$ generally correlates with $T_{\rm c}$.
     These results strongly indicate that AFM fluctuations play an important role in the appearance of SC in \fesex, as will be discussed below.   

    Figures \ref{fig:Fig3}(a)-\ref{fig:Fig3}(h) show the temperature dependence of 1/$T_1T$ at various pressures for $H||ab$ (gray circles) and $H||c$ (red circles).
   First let us discuss the temperature dependence of 1/$T_1T$ measured for $H||ab$, $(1/T_1T)_{ab}$. 
    In general, 1/$T_1T$ is related to the dynamical magnetic susceptibility as $1/T_1T\sim\gamma^{2}_{\rm N}k_{\rm B}\sum_{\mathbf{q}}|A(\mathbf{q})|^2\chi^{\prime\prime}(\mathbf{q}, \omega_{\rm N})/\omega_{\rm N}$, where $A(\mathbf{q})$ is the wave-vector $\mathbf{q}$ dependent form factor and $\chi^{\prime\prime}(\mathbf{q}, \omega_{\rm N})$ is the imaginary part of $\chi(\mathbf{q}, \omega_{\rm N})$ at the Larmor frequency $\omega_{\rm N}$  \cite{Moriya1963}. 
     Therefore, by comparing the temperature dependences between 1/$T_1T$ and $K$ which measures the $\mathbf{q}=0$ uniform magnetic susceptibility, one can obtain information on the temperature evolution of $\sum_{\mathbf{q}}\chi^{\prime\prime}(\mathbf{q}, \omega_{\rm N})$ with respect to that of $\chi^{\prime}(0, 0)$. 
    Above $\sim100$ K, $(1/T_1T)_{ab}$ shows a similar $T$ dependence as $K$ for all measured pressures. 
    On the other hand, below $\sim70$ K  the temperature dependence of  $(1/T_1T)_{ab}$ clearly deviates from that of $K$, although the enhancement of  $(1/T_1T)_{ab}$ becomes less pronounced at higher pressures. 
    This deviation of $1/T_1T$ at low $T$ therefore evidences the existence of AFM fluctuations with $\mathbf{q}\neq0$. 
 
   Below 0.5 GPa, with decreasing $T$, $(1/T_1T)_{ab}$ increases below $\sim$70 K  and starts to decrease around $T_{\rm c}$, making a broad maximum. 
   $T_{\rm c}$ for $H||ab$ are shown by black arrows. 
   The Curie-Weiss like behavior of $(1/T_1T)_{ab}$ above the maxima can be associated with  two dimensional AFM fluctuations \cite{Wiecki2018,Kuwayama2019}. 

   On the other hand, above 0.5 GPa, $(1/T_1T)_{ab}$ exhibits quite different temperature dependence in comparison with those observed at low pressures.
  Although $(1/T_1T)_{ab}$ is slightly enhanced below $\sim$ 70 K, indicating the existence of the AFM spin fluctuations,  we observe 1/$T_1T$ = constant, so-called Korringa behavior,  expected for Fermi-liquid state such as exchange enhanced metals \cite{Narath1968, Moriya1963} below  the temperature (defined as $T_{\rm FL}$) marked by blue arrows. 
    $T_{\rm FL}$ seems to increase from 40 K at $p$ = 0.9 GPa to 50 K at 1.70 GPa and then decreases to 30 K at 2.10 GPa. 
    The suppression of   $T_{\rm FL}$ at higher pressure may be due to the appearance of the AFM state under high pressures.
    It is important to point out that our NMR data do not indicate any quantum critical behavior due to nemacticty around 0.5 GPa.
    These results seem to be consistent with the recent resistivity studies under high magnetic fields \cite{Reiss2019} which reported  a lack of nematic quantum criticality and the presence of FL behavior in FeSe$_{0.89}$S$_{0.11}$ under pressure.  
    It is also worth to mention that no signatures of an AFM order were observed in 1/$T_1T$ as well as the NMR spectra,  in contrast to the recent $\mu$SR report on  FeSe$_{0.89}$S$_{0.11}$ under pressure \cite{Holenstein2019}. 
   It is not clear at present the reason why the AFM state reported by the $\mu$SR measurements is not detected by our NMR and resistivity measurements. Other experiments such as neutron diffraction measurements are highly required to elucidate the issue.

 \begin{figure}[tb]
\centering
\includegraphics[width=\columnwidth]{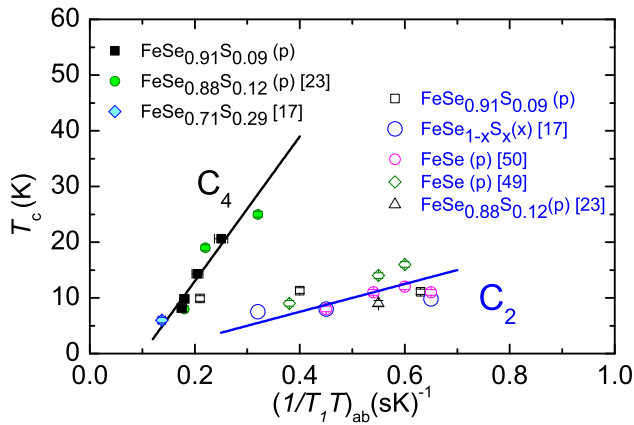} 
\caption{ Plot of $T_{\rm c}$ at zero field versus maximum values of $(1/T_1T)_{ab}$. 
     For $p$ $\leq$ 0.5 GPa, the values of $(1/T_1T)_{ab}$ are taken at the peak positions nearly just above $T_{\rm c}$. 
    Above $p$ =  0.5 GPa,   the constant values of $(1/T_1T)_{ab}$ below $T_{\rm FL}$ were used. 
     The solid and open squares are data from the present work. 
    The values for FeSe under $p$ were taken from Imai {\it et~al.} \cite{Imai2009} and Wiecki {\it et~al.} \cite{Wiecki2017}; for \fesex~under ambient $p$ from Wiecki {\it et~al.} \cite{Wiecki2018}; for FeSe$_{0.88}$S$_{0.12}$ under $p$ from Kuwayama {\it et~al} \cite{Kuwayama2019}. 
The black and blue lines show linear relations for AFM fluctuations with $C_4$ and $C_2$ symmetry, respectively.
}
\label{fig:Fig4}
\end{figure}   

  Our results indicate that the nature of AFM fluctuations changes below and above 0.5 GPa in \feses. 
  According to Kuwayama {\it et~al}. \cite{Kuwayama2019}, AFM fluctuations with different $\mathbf{q}$ vectors may be responsible for the  two distinct SC domes.
   Therefore, it is important to reveal the nature of the AFM fluctuations in the different pressure ranges.
  Based on previous NMR studies on Fe pnictides \cite{KitagawaSrFe2As2, Kitagawa2010, FukazawaKBaFe2As2} and related materials \cite{Furukawa2014,Pandey2013, Ding2016}, the ratio $R\equiv(1/T_1T)_{ab}/(1/T_1T)_{c}$  provides valuable information on \textbf{q} of the spin fluctuations. 
   In the case of isotropic spin fluctuations, $R$ = 1.5 is expected for stripe-type [\textbf{q}= ($\pi$,0) or (0,$\pi$)] fluctuations whereas $R$ = 0.5 for N\'eel type [\textbf{q}= ($\pi$,$\pi$)] fluctuations \cite{Kitagawa2010}. 
   Therefore, to determine the $p$ and $T$ dependence of $R$, we have measured 1/$T_1T$ at several pressures for  $H||c$ (shown by red circles in Fig.~\ref{fig:Fig3}). 
     As plotted in the inset of each panel of Fig.~\ref{fig:Fig3}, $R$ is $\sim$1 at temperatures above 200 K and increases to $R$ $\sim$1.5 at low temperatures below 100K throughout all measured pressures, although the data are slightly scattered, especially for 0.5 GPa. 
   It is important to note that $R$ never decreases down to 0.5 at any pressures.  
   Thus, one can conclude that the AFM fluctuations are characterized to be stripe-type and do not change in the lower and higher SC domes. 

    What then is the difference in AFM fluctuations between the SC1 and SC2 domes?
    One of the important changes in the character of AFM fluctuations is the presence or absence of nematic order, as has been discussed previously \cite{Fernandes2012,Fernandes2014}. 
    Below 0.5 GPa, the SC state arises from the nematic phase with $C_2$ symmetry. 
    In this case, the amplitude of AFM fluctuations with  \textbf{q$_{\rm x}$} = ($\pi$,0) and \textbf{q$_{\rm y}$} = (0,$\pi$) must be inequivalent. 
    On the other hand, since SC appears from the tetragonal phase above 0.5 GPa, the magnetic fluctuations with \textbf{q$_{\rm x}$} and \textbf{q$_{\rm y}$} are degenerate due to the $C_4$ symmetry. 

    In order to see how the relationship between SC and stripe-type AFM fluctuations changes with the symmetry, we plotted  the $T_{\rm c}$ at zero field versus the maximum value of $(1/T_1T)_{ab}$ below 100 K in Fig.~\ref{fig:Fig4}, together with data available from the literature.   
     When SC emerged from the nematic state with decreasing temperature, as in the case of FeSe for $p<$ 1.5 GPa \cite{Imai2009, Wiecki2017}, FeSe$_{0.88}$S$_{0.12}$ at ambient $p$ \cite{Kuwayama2019} and \fesex~for $x<$ 0.17 \cite{Wiecki2018} at ambient $p$, the AFM fluctuations are labeled as C$_2$. 
    When SC emerged in the tetragonal phase for FeSe$_{0.71}$S$_{0.29}$ \cite{Wiecki2018} at ambient $p$ and FeSe$_{0.88}$S$_{0.12}$ for $p>$ 0.5 GPa \cite{Kuwayama2019}, the AFM fluctuations are labeled as $C_4$. 
    This plot shows two different correlations between $T_{\rm c}$  and stripe-type AFM fluctuations with and without nematic order, indicating that the correlations hold for the whole FeSe$_{1-x}$S$_x$ system.
    When nematic order is absent, a clear and strong correlation between $T_{\rm c}$ and the stripe-type AFM fluctuations with $C_4$ symmetry exists as represented by the straight black line. 
    In contrast, when nematic order is present, $T_{\rm c}$ weakly depends on  the  stripe-type AFM fluctuations with $C_2$ symmetry, as represented by the blue line with a slope about 5 times smaller than that of the black line. 
   These results indicate that the AFM fluctuations with $C_4$ symmetry are more effective in enhancing the superconducting transition in the \fesex~system.

   In conclusion, by NMR and resistivity measurements under pressure, we have established the $p-T$ phase diagram of  \feses~exhibiting a nematic quantum phase transition around 0.5 GPa, two SC domes and an AFM phase above $\sim$3.3 GPa.  
    The AFM fluctuations evolve from non-Fermi liquid (Curie-Weiss like behavior of 1/$T_1T$) to a Fermi liquid behavior ($1/T_1T$=constant behavior) across the nematic QPT. 
     The stripe-type wavevector for the AFM fluctuations is revealed to be unchanged in the two SC domes, but the symmetry in the fluctuations is raised from $C_2$ to $C_4$ across the nematic QPT. 
  Although both AFM fluctuations are found to be correlated with $T_{\rm c}$ in \fesex~under pressure, our results clearly show that $T_{\rm c}$ is more sensitive to AFM fluctuations with $C_4$ symmetry than those with $C_2$ symmetry.

We thank Qing-Ping Ding and Elena Gati for helpful discussions.
The research was supported by the U.S. Department of Energy (DOE), Office of Basic Energy Sciences, Division of Materials Sciences and Engineering. Ames Laboratory is operated for the U.S. DOE by Iowa State University under Contract No.~DE-AC02-07CH11358.


\newpage

\clearpage

\section{Supplementary Material}

\setcounter{figure}{0}

\section{ac-susceptibility Measurements}

$In$ $situ$ ac-susceptibility ($\chi_{ac}$) using the NMR coil was measured to determine the superconducting transition temperature, ($T_{c,\chi_{ac}}$). The NMR coil tank circuit resonance frequency ($f$) is a measure of $\chi_{ac}$ with the relation, $f=\frac{1}{2\pi\sqrt{L_0(1+\chi_{ac})C}}$ where $ L_0$ and C are the inductance and capacitance in the circuit respectively. In the superconducting state, $\chi_{ac}$ decreases due to Meissner effect, there by increasing $f$.  The temperature dependence of $f$ was measured in zero field and under a magnetic field of 7.4089 T in the $ab$-plane direction as shown in Figs.~\ref{fig:Fig1}(a) and \ref{fig:Fig1}(b) respectively. $T_{\rm c,\chi_{ac}}$ was determined as the crossing points of the straight lines, as marked by downward arrows.

\section{Knight shift as a function of pressure}

 Figures~\ref{Fig6}(a) and \ref{Fig6}(b) show the $p$ dependence of Knight shift ($K$) at various temperatures for $H\|ab$ ($K_{ab}$) and $H\| c$ ($K_{c}$) directions, respectively. 
 $K_{\rm ab}$ decreases slightly with $p$, while the magnitude of decrements increases with temperature. 
 $K_{\rm c}$ is nearly independent of $p$ below 80 K, although $K$  shows monotonic decrease with $p$ for higher temperatures. 
Figure \ref{Fig6}(c) shows the pressure dependence of the average value of Knight shift derived as $K_{\rm ave}=\frac{1}{3}(2K_{ab}+K_{c})$ which also exhibits a nearly pressure independent behavior at low temperatures.  
 Since Knight shift is proportional to the density of states at the Fermi energy, $N(E_{\rm F})$, these results indicate that 
$N(E_{\rm F})$ is nearly independent of $p$, especially at low temperatures.  
 In contrast to conventional BCS superconductors, $N(E_{\rm F})$ does not show correlation with $T_{\rm c}$  which shows a double dome structure with $p$, suggesting the importance of AFM fluctuations in this system. 


\begin{figure}[tb]
\includegraphics[width=\columnwidth]{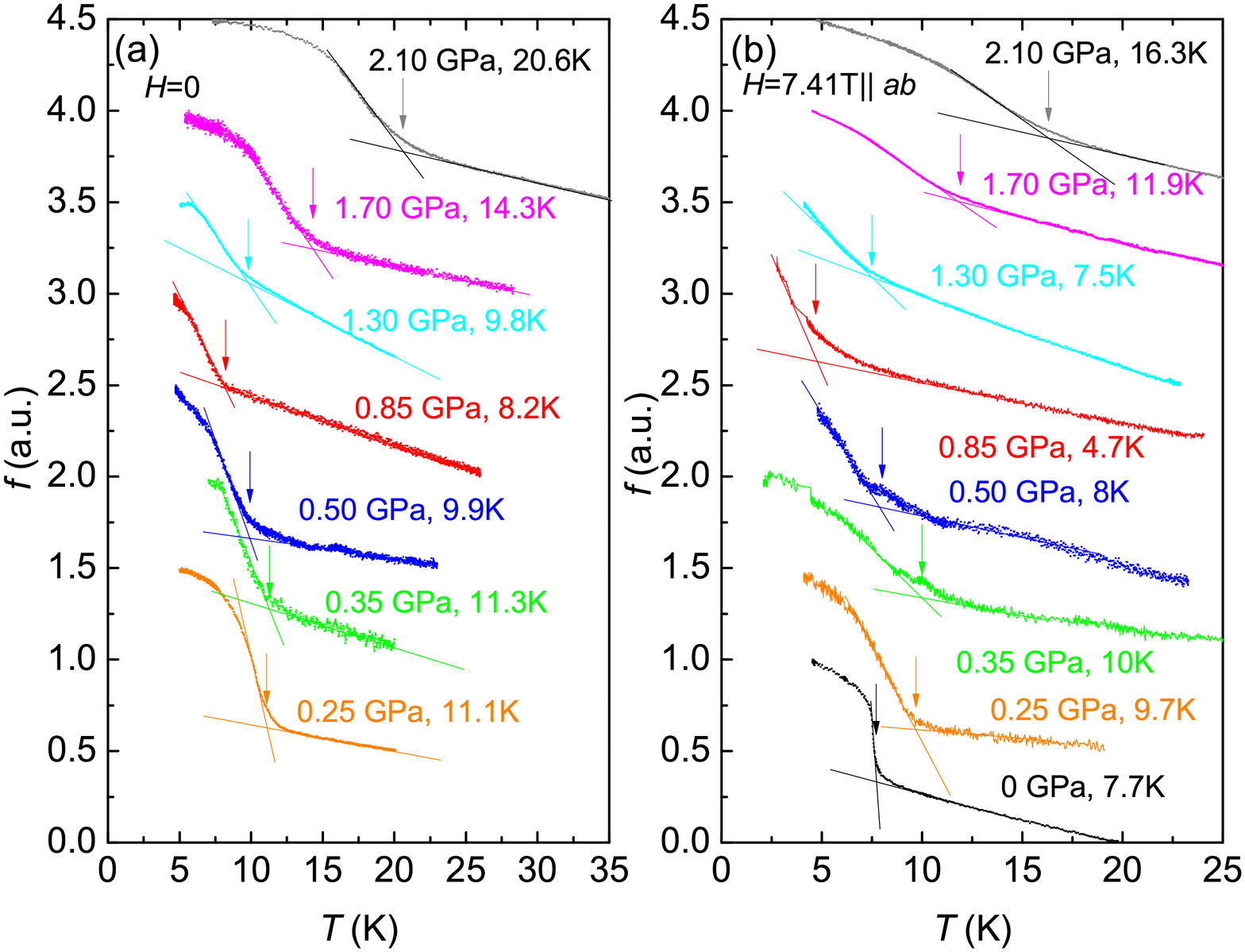} 
\caption{ (a), (b) NMR coil tank circuit resonance frequency ($f$) as a function of temperature ($T$) with pressure ($p$) as an implicit parameter at $H=0$  and  $H\|ab$, respectively. 
 Downward arrows represent the superconducting transition temperature ($T_{\rm c}$), which indicate the characteristic temperature at which $\chi_{ac}$ starts to decrease due to superconductivity. Data sets are offset for clarity.} 
\label{fig:Fig1}
\end{figure}

\begin{figure}[tb]
\includegraphics[width=\columnwidth]{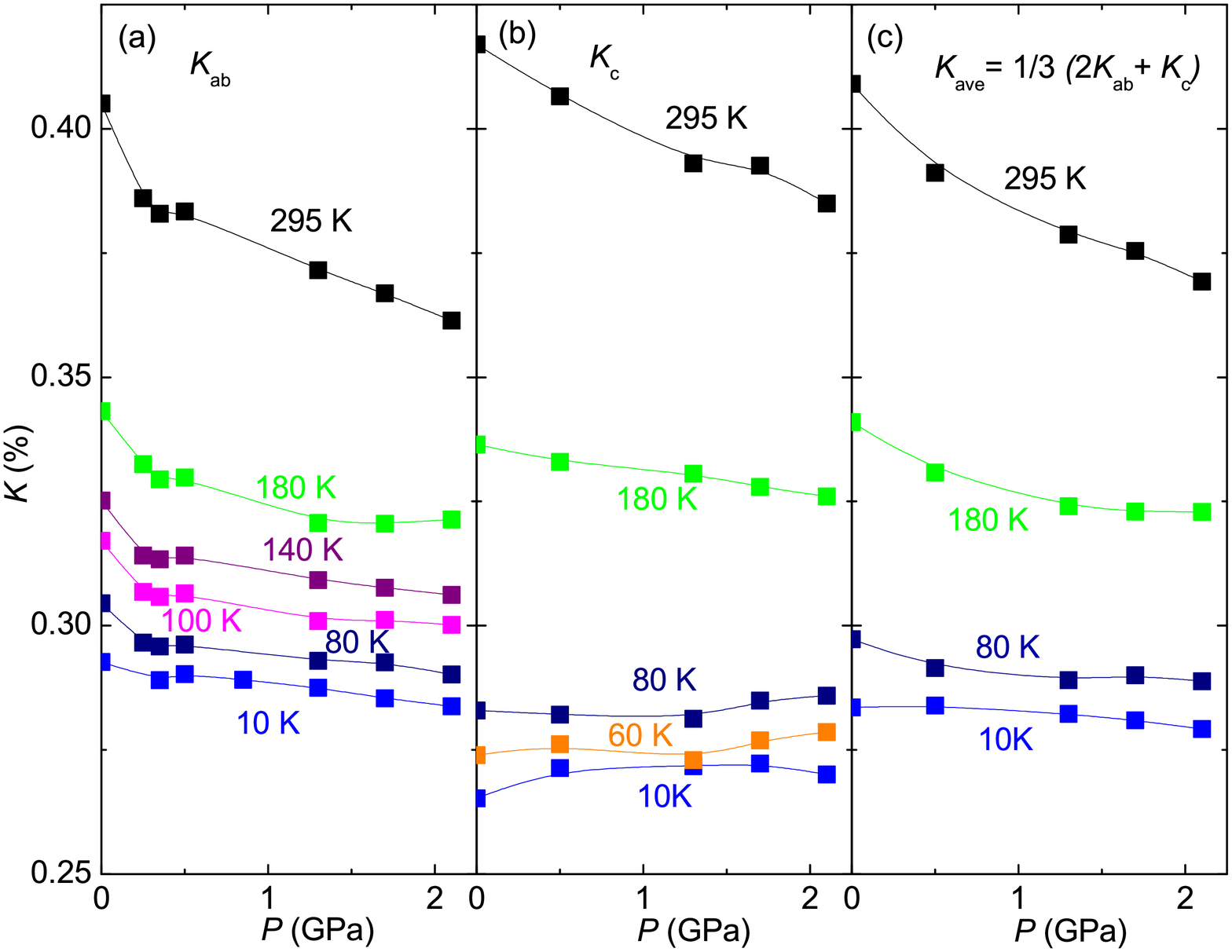} 
\caption{Pressure dependence of Knight Shift ($K$) at several temperatures.  (a) $K_{ab}$,  (b) $K_{c}$ and (c) $K_{\rm ave}=\frac{1}{3}(2K_{ab}+K_{c})$.}
\label{Fig6}
\end{figure}

\section{Resistivity Measurements}
The in-plane, ac resistance measurements, with current flow along the $ab$ plane, under pressure were performed in a Quantum Design Physical Property Measurement System (PPMS) using a 1 mA excitation with frequency of 17 Hz, with a cooling rate of 0.25 K/min. A standard, linear four-contact configuration was used. Contacts were made by spot welding 25 $\mu$m Au wires on top of the sample. The magnetic field was applied along the $c$ axis. A modified Bridgman Anvil Cell (mBAC) \cite{Colombier2007} was used to apply pressure up to 4.73 GPa. Pressure values at low temperature were inferred from the $T_\textrm{c}(p)$ of lead \cite{Bireckoven1988}. Hydrostatic conditions were achieved by using a 1:1 mixture of iso-pentane:n-pentane as the pressure medium, which solidifies at $\sim$ 6.5 GPa at room temperature \cite{Torikachvili2015}.

	\begin{figure}
		\begin{center}
			\includegraphics[width=\columnwidth]{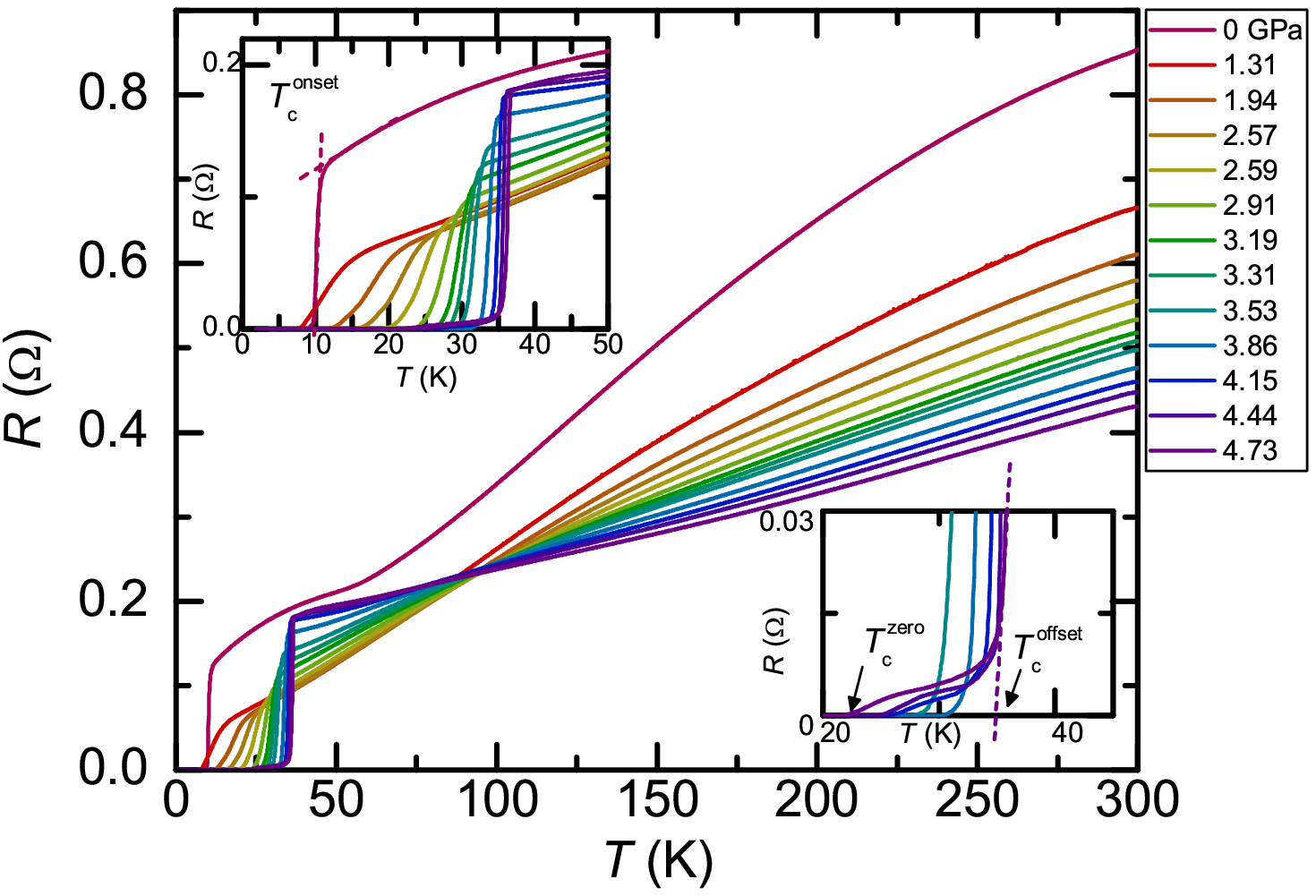} 
			\caption{Evolution of temperature dependent resistance for FeSe$_{0.91}$S$_{0.09}$ under pressure up to 4.73 GPa. Upper inset: Blow up of the resistance data at low temperature showing the superconducting transition. Bottom inset: Blow up of the resistance data showing that at our highest pressures ($p\geq$ 4.15 GPa), the superconducting transition has long tail. Superconducting transition temperature using onset ($T_\mathrm{c}^\mathrm{onset}$), offset ($T_\mathrm{c}^\mathrm{offset}$) and zero-resistance ($T_\mathrm{c}^\mathrm{zero}$) criteria are indicated by arrows in the figure.}
			\label{RT_4th}
		\end{center}
	\end{figure}
	
Figure \ref{RT_4th} shows resistance ($R$) as a function of $T$ for various $p$. At ambient pressure, the superconducting transition is comparatively sharp and broadens under initial pressurizing for $p\le$ 1.94 GPa. Further increasing pressure sharpens the superconducting transition again. At even higher pressures ($p\geq$ 4.15 GPa), the superconducting transition has long tail as shown in the bottom inset. The superconducting transition temperatures using onset ($T_\mathrm{c}^\mathrm{onset}$), offset ($T_\mathrm{c}^\mathrm{offset}$) and zero-resistance ($T_\mathrm{c}^\mathrm{zero}$) criteria are indicated by arrows in the figure.
	
	\begin{figure}
		\begin{center}
			\includegraphics[width=\columnwidth]{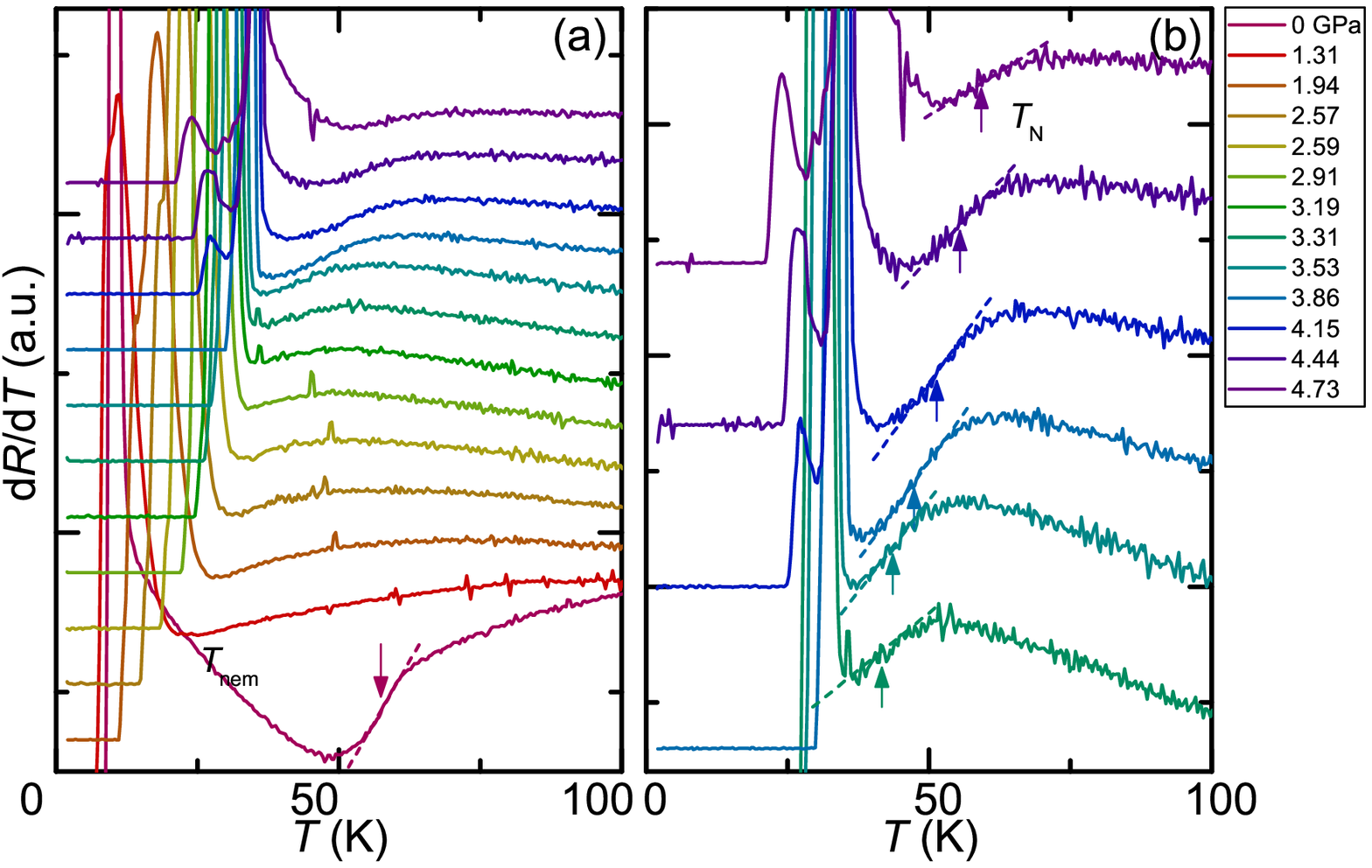} 
			\caption{(a) Temperature derivative of the resistance, d$R$/d$T$, showing the evolution of the structural transition $T_\mathrm{nem}$ and magnetic transition $T_\mathrm N$. $T_\mathrm{nem}$ at ambient pressure is shown by the downward arrow. (b) Blow up of d$R$/d$T$ at our highest pressures. $T_\mathrm{N}$ values are marked by the upward arrows. Data sets are offset for clarity.}
			\label{dRdT}
		\end{center}
	\end{figure}
	
Figure \ref{dRdT} shows temperature derivative, d$R$/d$T$, as a function of $T$ at the measured pressures. At ambient pressure, a step-like anomaly (shown as the color-coded dashed line in Fig. \ref{dRdT} (a)) associated with structural transition, $T_\mathrm{nem}$, is seen around 60 K, which is suppressed with the first applied pressure at 1.31 GPa. At higher pressures, ($p\gtrsim$ 3.31 GPa), more subtle step-like anomalies (shown as color-coded dashed lines in Fig. \ref{dRdT} (b)) are observed before the rapid increase in d$R$/d$T$ due to the resistance drop from superconductivity. These step-like anomalies are associated with the magnetic transition $T_\mathrm N$ at high pressures. The $T_\mathrm{nem}$ and $T_\mathrm N$ values are determined as the middle point of the step-like anomalies [marked by arrows in Figs. \ref{dRdT} (a) and (b)] with error bars taken from the corresponding temperatures where d$R$/d$T$ deviates from the dashed lines. As shown in Fig. \ref{dRdT} (b), at 3.31 GPa, the step-like anomaly for $T_\mathrm N$ is already close to the superconducting transition and hence becomes hard to resolve for low pressures.
	
	\begin{figure}
		\begin{center}
			\includegraphics[width=\columnwidth]{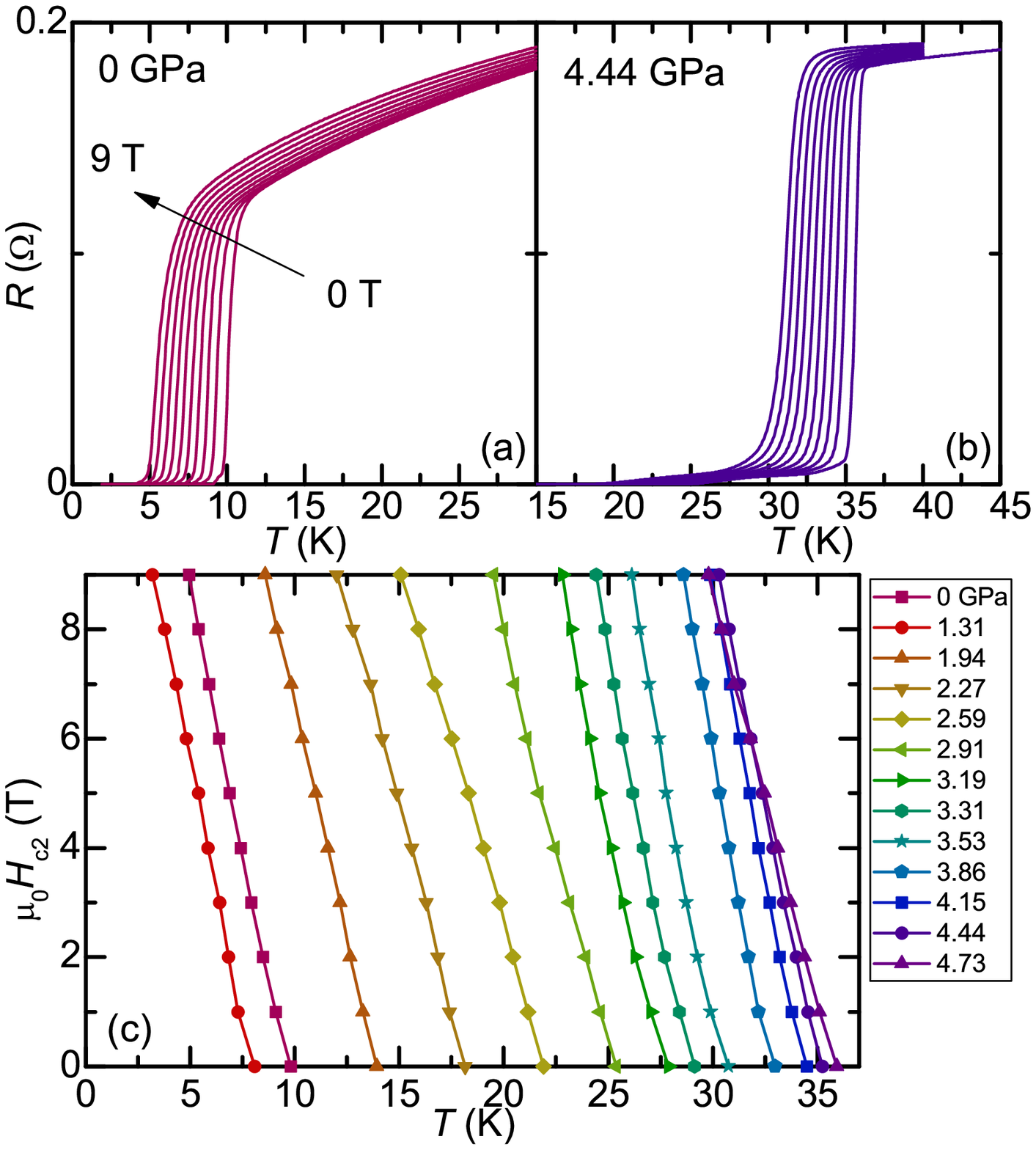} 
			\caption{(a), (b) Temperature dependent resistance data under magnetic fields up to 9 T under representative pressures 0 GPa and 4.44 GPa. (c) Temperature dependence of the upper superconducting critical field $H_\mathrm {c2}$ under various pressures up to 4.73 GPa using $T_\mathrm{c}^\mathrm{offset}$.}
			\label{Hc2}
		\end{center}
	\end{figure}
	
Temperature dependent resistance under magnetic field up to 9 T was measured for various pressures. Figure \ref{Hc2} (a) and (b) present such $R(T)$ data for representative pressures, 0 GPa and 4.44 GPa. Furthermore, the temperature dependent upper critical field $H_\mathrm{c2}$ for various pressures are obtained using $T_\mathrm{c}^\mathrm{offset}$ (as it represents the dominating superconducting transition at all pressures and traces $T_\mathrm {c}(p)$ similarly as $T_\mathrm{c}^\mathrm{onset}$)and presented in Fig. \ref{Hc2}(c).

	\begin{figure}
		\begin{center}
			\includegraphics[width=\columnwidth]{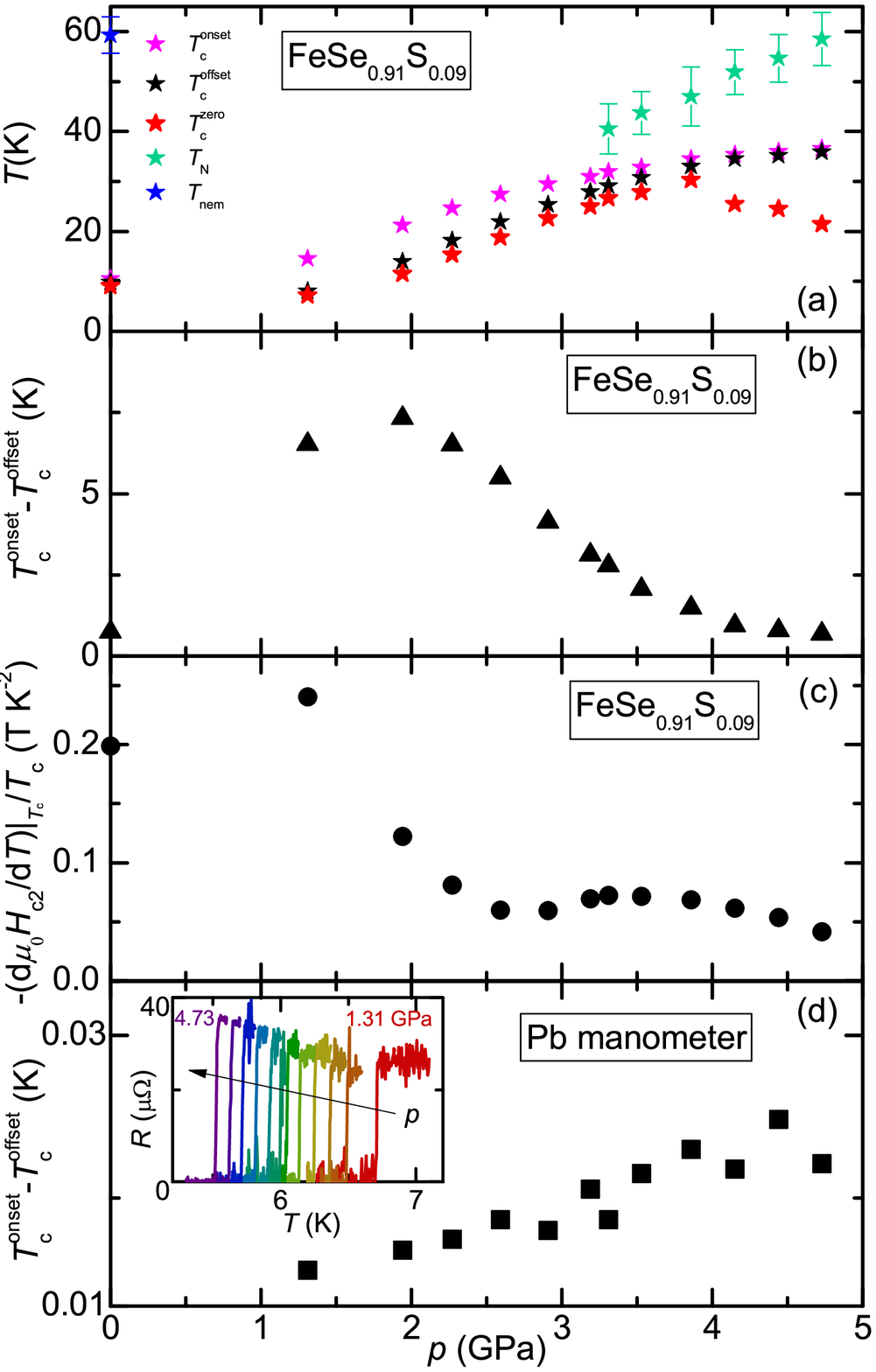} 
			\caption{(a) Pressure-temperature phase diagram of FeSe$_{0.91}$S$_{0.09}$ as determined from resistance measurements including structural transition $T_\mathrm{nem}$ (blue), magnetic transition $T_\mathrm{N}$ (green) as well as superconducting transition temperature using different criteria, $T_\mathrm{c}^\mathrm{onset}$ (magenta), $T_\mathrm{c}^\mathrm{offset}$ (black) and $T_\mathrm{c}^\mathrm{zero}$ (red). (b) Pressure dependence of the superconducting transition width defined as $T_\mathrm{c}^\mathrm{onset}$-$T_\mathrm{c}^\mathrm{offset}$. (c) Pressure dependence of the normalized upper critical field slope, -(1/$T_\mathrm c$)($d\mu_oH_{\mathrm c2}$/$dT$)$|_{T_\mathrm c}$, using offset superconducting transition temperature $T_\mathrm{c}^\mathrm{offset}$. (d) Pressure dependence of the superconducting transition width ($T_\mathrm{c}^\mathrm{onset}$-$T_\mathrm{c}^\mathrm{offset}$) of the Pb manometer in mBAC. Inset: temperature dependent resistance of Pb manometer under various pressures showing superconducting transition.}
			\label{phasediagram}
		\end{center}
	\end{figure}

The values of the superconducting transition temperature, $T_\mathrm c$ (using different criteria), structural transition temperature, $T_\mathrm{nem}$, as well as magnetic transition temperature, $T_\mathrm N$, determined from resistance measurements using the criteria outlined in Figs. \ref{RT_4th} and \ref{dRdT} are summarized and presented in a $p-T$ phase diagram in Fig. \ref{phasediagram}(a). As shown in the figure, $T_\mathrm{nem}$ is completely suppressed before the first applied pressure (1.31 GPa) in mBAC ($T_\mathrm{nem}$ is suppressed by $\sim$ 0.5 GPa as evidenced by the NMR measurements shown in the main text). As for the superconducting transition temperature, $T_\mathrm{c}^\mathrm{onset}$ and $T_\mathrm{c}^\mathrm{offset}$ show similar pressure dependences and monotonically increase with pressure above 1.31 GPa and slowly saturate at our highest pressures. $T_\mathrm{c}^\mathrm{zero}(p)$ deviates from $T_\mathrm{c}^\mathrm{onset}(p)$ and $T_\mathrm{c}^\mathrm{offset}(p)$ above 3.86 GPa by showing decrease with increasing pressure. This is due to the long-tail resistive behavior at high pressures as pointed out in Fig. \ref{RT_4th}. At $p\gtrsim$ 3.31 GPa, magnetic transitions are resolved from $R(T)$ data, of which the temperature, $T_\mathrm{N}$, increases with increasing pressure.

Pressure dependence of the superconducting transition width, defined as $T_\mathrm{c}^\mathrm{onset}$-$T_\mathrm{c}^\mathrm{offset}$, is presented in Fig. \ref{phasediagram} (b). $T_\mathrm{c}^\mathrm{onset}$-$T_\mathrm{c}^\mathrm{offset}$ shows a non-monotonic dependence on $p$ manifesting a maximum at $\sim$ 1.94 GPa. It is noteworthy that the transition width at the highest pressure (0.68 K) is even smaller than that at ambient pressure (0.74 K) outside of pressure cell. It is known that the broadening of the superconducting transition inside pressure cell can be caused by the pressure inhomogeneity. The pressure inhomogeneity in this study is estimated by the superconducting transition width of Pb manometer in mBAC as shown Fig. \ref{phasediagram} (d). $T_\mathrm{c}^\mathrm{onset}$-$T_\mathrm{c}^\mathrm{offset}$ of Pb shows an overall increase with $p$. This suggests that the non-monotonic dependence of transition width of specimen can not be simply explained just by the pressure inhomogeneity.

Upper critical field $H_\mathrm{c2}$ is further analyzed by calculating the the normalized slope of $H_\mathrm{c2}(T)$. Generally speaking, the slope of $H_\mathrm{c2}(T)$ normalized by $T_\mathrm{c}$, is related to the Fermi velocity and superconducting gap of the system \cite{Kogan2012}. In the clean limit, for a single-band,
\begin{equation}
-(1/T_\mathrm c)(d\mu_oH_{\mathrm c2}/dT)|_{T_\mathrm c} \propto 1/v_F^2,
\label{eq:Hc2}
\end{equation}
where $v_F$ is the Fermi velocity. A change in the normalized slope of $H_\mathrm{c2}$ can indicate or be attributed to changes in the Fermi surface, the superconducting gap structure, or the pairing mechanism \cite{Kogan2012, Kogan2014, Taufour2014, Kaluarachchi2016}. Figure \ref{phasediagram} (c) presents the pressure dependence of the normalized $H_\mathrm{c2}$ slope. As shown in the figure, $-(1/T_\mathrm c)(d\mu_oH_{\mathrm c2}/dT)|_{T_\mathrm c}$ shows a very non-monotonic dependence on $p$. At low pressures, the changes of $-(1/T_\mathrm c)(d\mu_oH_{\mathrm c2}/dT)|_{T_\mathrm c}$ from 0 GPa to 1.31 GPa could be related to a possible magnetic phase as observed in both lower ($x$ = 0.043) and higher ($x$ = 0.096) substituted FeSe$_{1-x}$S$_x$ systems \cite{Xiang2017PRB}.
 At $\sim$ 3.31 GPa, a local maximum in $-(1/T_\mathrm c)(d\mu_oH_{\mathrm c2}/dT)|_{T_\mathrm c}$ is also observed, which could be related to Fermi surface reconstruction due to the emerging magnetic phase at $p\gtrsim$ 3.31 GPa [Fig. \ref{phasediagram} (a)].


\begin{thebibliography}{10}


\bibitem{Kamihara2008} Y. Kamihara, T. Watanabe, M. Hirano, and H. Hosono, J. Am. Chem. Soc. {\bf 130}, 3296 (2008).
\bibitem{Johnston2010} D. C. Johnston, Adv.  Phys. {\bf 59}, 803 (2010).
\bibitem{Canfield2010} P. C. Canfield and S. L. Bud'ko, Annu. Rev. Condens. Matter Phys. {\bf 1}, 27 (2010).
\bibitem{Stewart2011} G. R. Stewart, Rev. Mod. Phys. {\bf 83}, 1589 (2011).
\bibitem{Scalipino2013} D. J. Scalapino, Rev. Mod. Phys. {\bf 84}, 1383 (2012).
\bibitem{Bohmer2018} A. E. B\"{o}hmer and A. Kreisel, J. Phys.: Condens. Matter {\bf 30}, 023001 (2018).

\bibitem{Hsu2008} F.-C. Hsu, J.-Y. Luo, K.-W. Yeh, T.-K. Chen, T.-W. Huang, P. M. Wu, Y.-C. Lee, Y.-L. Huang, Y.-Y. Chu, D.-C. Yan, and M.-K. Wu, Proc. Natl. Acad. Sci. U.S.A. {\bf 105}, 14262 (2008). 
\bibitem{McQueen2009} T. M. McQueen, A. J. Williams, P. W. Stephens, J. Tao, Y. Zhu, V. Ksenofontov, F. Casper, C. Felser, and R. J. Cava,
Phys. Rev. Lett. {\bf 103}, 057002 (2009).

\bibitem{Hosoi2016}S. Hosoi, K. Matsuura, K. Ishida, H. Wang, Y. Mizukami, T. Watashige, S. Kasahara, Y. Matsuda, and T. Shibauchi, Proc. Natl. Acad. Sci. USA {\bf 113}, 8139 (2016).
\bibitem{Abdel2015}Mahmoud Abdel-Hafiez, Yuan-Yuan Zhang, Zi-Yu Cao, Chun-Gang Duan, G. Karapetrov, V. M. Pudalov, V. A. Vlasenko, A. V. Sadakov, D. A. Knyazev, T. A. Romanova, D. A. Chareev, O. S. Volkova, A. N. Vasiliev, and Xiao-Jia Chen,
Phys. Rev. B {\bf 91}, 165109 (2015).
\bibitem{Watson2015}
M. D. Watson, T. K. Kim, A. A. Haghighirad, S. F. Blake, N. R. Davies, M. Hoesch, T. Wolf, and A. I. Coldea
Phys. Rev. B {\bf  92}, 121108(R) (2015).
\bibitem{Reiss2017} P. Reiss, M. D. Watson, T. K. Kim, A. A. Haghighirad, D. N. Woodruff, M. Bruma, S. J. Clarke, and A. I. Coldea,
Phys. Rev. B {\bf 96}, 121103(R) (2017).
\bibitem{Lai2015} X. Lai, H. Zhang, Y. Wang, X. Wang, X. Zhang, J. Lin, and F. Huang, J. Am. Chem. Soc., {\bf 137} 10148 (2015).



\bibitem{Hanaguri2018}T. Hanaguri, K. Iwaya, Y. Kohsaka, T. Machida, T. Watashige, S. Kasahara, T. Shibauchi, and Y. Matsuda, Sci. Adv. {\bf 4}, eaar6419 (2018).
\bibitem{Sato2018}Y. Sato, S. Kasahara, T. Taniguchi, X. Z. Xing, Y. Kasahara,Y. Tokiwa, T. Shibauchi, and Y. Matsuda, Proc. Natl. Acad. Sci. U.S.A. {\bf 115}, 1227 (2018).
\bibitem{Coldea2019}A. I. Coldea, S. F. Blake, S. Kashara, A. A. Haghighirad, M. D. Watso, W. Knafo, E. S. Choi, A. McCollam, P. Reiss, T. Yamashita, M. Bruma, S. C. Speller, Y. Matsudea, T. Wolf, T. Shibauchi, A. J. Schofield,  npj Quant Mater {\bf 4}, 2 (2019).

\bibitem{Wiecki2018}P. Wiecki, K. Rana, A. E. B\"ohmer, Y. Lee, S. L. Bud'ko, P. C. Canfield, and Y. Furukawa, Phys. Rev. B {\bf 98} 020507(R) (2018).
\bibitem{Beak2020} S.-H. Baek, J. M. Ok, J. S. Kim, S. Aswartham, I. Morozov, D. Chareev, T. Urata, K. Tanigaki, Y. Tanabe, B. B\"uchner, D. V. Efremov, 
arXiv:2001.02079.



\bibitem{Matsuura2017} K. Matsuura, Y. Mizukami, Y. Arai, Y. Sugimura, N. Maejima, A. Machida, T. Watanuki, T. Fukuda, T. Yajima, Z. Hiroi, K., Y. Yip, Y. C. Chan, Q. Niu, S. Hosoi, K. Ishida, K. Mukasa, T. Watashige, S. Kasahara, J.-G. Cheng, S. K. Goh, Y. Matsuda, Y. Uwatoko, and T. Shibauchi, Nat. Commun. {\bf 8}, 1143 (2017).
\bibitem{Xiang2017} L. Xiang, U. S. Kaluarachchi, A. E. B\"{o}hmer, V. Taufour, M. A. Tanatar, R. Prozorov, S. L. Bud'ko, and P. C. Canfield,
Phys. Rev. B {\bf 96}, 024511 (2017).

\bibitem{Holenstein2019} S. Holenstein, J. Stahl, Z. Shermadini, G. Simutis, V. Grinenko, D. A. Chareev, R. Khasanov, J.-C. Orain, A. Amato, H.-H. Klauss, E. Morenzoni, D. Johrendt, and H. Luetkens, Phys. Rev. Lett. {\bf 123}, 147001 (2019).

\bibitem{Reiss2019} P. Reiss, D. Graf, A. A. Haghighirad, W. Knafo, L. Drigo, M. Bristow, A. J. Schofield, A.I. Coldea, Nat. Phys., s41567-019-0694-2 (2019). 
\bibitem{Kuwayama2019}T. Kuwayama, K. Matsuura, Y. Mizukmami, S. Kasahara, Y. Matsuda, T. Shibauchi, Y. Uwatoko, and N. Fujiwara, J. Phys. Soc. Jpn. {\bf 88}, 033703 (2019).
\bibitem{supp} See supplemental material for the experimental details, the ac susceptibility measurements using ${\it in~situ}$ NMR coil under $H$ = 0 and 7.4089 T, the pressure dependence of Knight shift at several temperatures, the resistivity data under pressure and magnetic field, and the pressure dependence of $H_{\rm c2}$ determined by the resistivity measurements,  which includes Refs. \cite{Colombier2007, Bireckoven1988,Torikachvili2015,Kogan2012,Kogan2014,Taufour2014,Kaluarachchi2016,Xiang2017PRB}.

\bibitem{Colombier2007} E. Colombier and D. Braithwaite, Review of Scientific Instruments {\bf78}, 093903 (2007).
\bibitem{Bireckoven1988} B. Bireckoven and J. Wittig, Journal of Physics E: Scientific Instruments {\bf21}, 841 (1988).
\bibitem{Torikachvili2015} M. S. Torikachvili, S. K. Kim, E. Colombier, S. L. Bud'ko, and P. C. Canfield, Rev. Sci. Instrum. {\bf86}, 123904 (2015).

\bibitem{Kogan2012}V. G. Kogan and R. Prozorov, Rep. Prog. Phys. {\bf75}, 114502 (2012).
\bibitem{Kogan2014}V. G. Kogan and R. Prozorov, Phys. Rev. B {\bf90}, 180502(R) (2014).
\bibitem{Taufour2014} V. Taufour, N. Foroozani, M. A. Tanatar, J. Lim, U. Kaluarachchi, S. K. Kim, Y. Liu, T. A. Lograsso, V.G. Kogan, R. Prozorov, S. L. Bud'ko, J. S. Schilling and P. C. Canfield, Phys. Rev. B {\bf89}, 220509(R)  (2014). 
\bibitem{Kaluarachchi2016} U. S. Kaluarachchi, V. Taufour, A. E. B\"ohmer, M. A. Tanatar, S. L. Bud'ko, V. G. Kogan, R. Prozorov and P.C. Canfield, Phys. Rev. B {\bf93}, 064503 (2016). 
\bibitem{Xiang2017PRB}L. Xiang, U. S. Kaluarachchi, A. E. B\"ohmer, V. Taufour, M. A. Tanatar, R. Prozorov, S. L. Bud'ko and P.C. Canfield, Phys. Rev. B {\bf96}, 024511 (2017).


\bibitem{Bohmer2016} A. E. B\"ohmer, V. Taufour, W. E. Straszheim, T. Wolf, and P. C. Canfield, Phys. Rev. B {\bf 94}, 024526 (2016).
\bibitem{NMR}      $^{\rm 77}$Se NMR spectra were measured at a fixed magnetic field $H$ = 7.4089 T using Fast Fourier Transform method. 
     1/$T_1$ was measured using a saturation method and determined by single exponential fitting for measured nuclear magnetization recovery behavior.

\bibitem{Fukazawa2007}
H. Fukazawa, N. Yamatoji, Y. Kohori, C. Terakura, N. Takeshita, Y. Tokura and H. Takagi, Rev. Sci. Instrum. {\bf 78}, 015106 (2007).
\bibitem{Reyes1992}
A. P. Reyes, E. T. Ahrens, R. H. Heffner, P. C. Hammel, and J. D. Thompson, Rev. Sci. Instrum. {\bf 63}, 3120 (1992).
\bibitem{K_details} See supplemental materials for more details of the pressure dependence of $K$ at several temperatures.   
\bibitem{K data}   It is worth to mention that the clear change of $N(E_{\rm F}$) due to a Lifshitz transition observed in quantum oscillation measurements at $x$ = 0.11 under $p$ $\sim$ 0.5 GPa \cite{Coldea2019} was not detected by our $K$ measurements in \feses.
 

\bibitem{Moriya1963} T. Moriya, J. Phys. Soc. Jpn. {\bf 18}, 516 (1963).
\bibitem{Narath1968} A. Narath and H. T. Weaver, Phys. Rev. {\bf 175}, 373 (1968).
\bibitem{KitagawaSrFe2As2} K. Kitagawa, N. Katayama, K. Ohgushi, and M. Takigawa, J. Phys. Soc. Jpn. {\bf 78}, 063706 (2009).
\bibitem{Kitagawa2010} S. Kitagawa, Y. Nakai, T. Iye, K. Ishida, Y. Kamihara, M. Hirano, and H. Hosono, Phys. Rev. B {\bf 81}, 212502 (2010).
\bibitem{FukazawaKBaFe2As2} M. Hirano, Y. Yamada, T. Saito, R. Nagashima, T. Konishi, T. Toriyama, Y. Ohta, H. Fukazawa, Y. Kohori, Y. Furukawa, K. Kihou, C.-H. Lee, A. Iyo and H. Eisaki, J. Phys. Soc. Jpn. {\bf 81}, 054704  (2012).
\bibitem{Furukawa2014} Y. Furukawa, B. Roy, S. Ran, S. L. Bud'ko, and P. C. Canfield, Phys. Rev. B {\bf 89}, 121109(R) (2014).
\bibitem{Pandey2013} A. Pandey, D. G. Quirinale, W. Jayasekara, A. Sapkota, M. G. Kim, R. S. Dhaka, Y. Lee, T. W. Heitmann, P. W. Stephens, V. Ogloblichev, A. Kreyssig, R. J. McQueeney, A. I. Goldman, A. Kaminski, B. N. Harmon, Y. Furukawa, and D. C. Johnston, Phys. Rev. B {\bf 88}, 014526 (2013).
\bibitem{Ding2016} Q.-P. Ding, P. Wiecki, V. K. Anand, N. S. Sangeetha, Y. Lee, D. C. Johnston, and Y. Furukawa, Phys. Rev. B {\bf 93}, 140502(R) (2016).
\bibitem{Fernandes2012} R. M. Fernandes, A. V. Chubukov, J. Knolle, I. Eremin, and J. Schmalian, Phys. Rev. B {\bf 85}, 024534 (2012).
\bibitem{Fernandes2014} R. M. Fernandes, A. V. Chubukov, and J. Schmalian, Nature Phys. {\bf 10}, 97 (2014). 
\bibitem{Imai2009}T. Imai, K. Ahilan, F. L. Ning, T. M. McQueen, and R. J. Cava, Phys. Rev. Lett. 102, 177005 (2009).
\bibitem{Wiecki2017}P. Wiecki, M. Nandi, A. E. B\"ohmer, S. L. Bud'ko, P. C. Canfield, and Y. Furukawa, Phys. Rev. B {\bf 96} 180502(R) (2017).



\end{thebibliography}

\begin{thebibliography}{10}
\bibitem{Colombier2007} E. Colombier and D. Braithwaite, Review of Scientific Instruments {\bf78}, 093903 (2007).
\bibitem{Bireckoven1988} B. Bireckoven and J. Wittig, Journal of Physics E: Scientific Instruments {\bf21}, 841 (1988).
\bibitem{Torikachvili2015} M. S. Torikachvili, S. K. Kim, E. Colombier, S. L. Bud’ko, and P. C. Canfield, Rev. Sci. Instrum. {\bf86}, 123904 (2015).
\bibitem{Kogan2012}V. G. Kogan and R. Prozorov, Rep. Prog. Phys. {\bf75}, 114502 (2012).
\bibitem{Kogan2014}V. G. Kogan and R. Prozorov, Phys. Rev. B {\bf90}, 180502 (2014).
\bibitem{Taufour2014} V. Taufour, N. Foroozani, M. A. Tanatar, J. Lim, U. Kaluarachchi, S. K. Kim, Y. Liu, T. A. Lograsso, V.G. Kogan, R. Prozorov, S. L. Bud'ko, J S. Schilling and P. C. Canfield, Phys. Rev. B {\bf89}, 220509 (2014). 
\bibitem{Kaluarachchi2016} U. S. Kaluarachchi, V. Taufour, A. E. B\"ohmer, M. A. Tanatar, S. L. Bud'ko, V. G. Kogan, R. Prozorov and P.C. Canfield, Phys. Rev. B {\bf93}, 064503 (2016). 
\bibitem{Xiang2017PRB}L. Xiang, U. S. Kaluarachchi, A. E. B\"ohmer, V. Taufour, M. A. Tanatar, R. Prozorov, S. L. Bud'ko and P.C. Canfield, Phys. Rev. B {\bf96}, 024511 (2017). 

\end{thebibliography}
\end{document}